\documentclass[showpacs,preprint,aps,tightenlines]{revtex4}%
\usepackage{amsfonts}
\usepackage{amsmath}
\usepackage{amssymb}
\usepackage{graphicx}%
\providecommand{\U}[1]{\protect\rule{.1in}{.1in}}

\begin{document}
\preprint{LM11677}
\title[ ]{Phase separation in imbalanced fermion superfluids beyond mean-field}
\author{J. Tempere}
\altaffiliation{Also at the Lyman Laboratory of Physics, Harvard University, Cambridge, MA 02138.}

\author{S. N. Klimin}
\altaffiliation{Permanent address: Department of Theoretical Physics, State University of
Moldova, str. A. Mateevici 60, MD-2009 Kishinev, Republic of Moldova.}

\author{J. T. Devreese}
\altaffiliation{Also at Technische Universiteit Eindhoven, P. B. 513, 5600 MB Eindhoven, The Netherlands.}

\affiliation{Theoretische Fysica van de Vaste Stoffen (TFVS), Universiteit Antwerpen,
B-2020 Antwerpen, Belgium}

\begin{abstract}
The density distributions of the two components of a trapped, ultracold Fermi
gas with population imbalance reveal the effect of imbalance on superfluid
pairing. We develop a path-integral derivation of the density, that takes into
account both fluctuations beyond mean-field and effects of nonzero
temperature. The calculated density profiles compare favorably to the measured
density profiles, and illustrate the necessity to incorporate both quantum
fluctuations and finite temperature effects. The temperature dependence of the
density profiles, especially near the superfluid-normal interface, allows
determining the temperature of the superfluid core in current experiments.

\end{abstract}
\volumeyear{ }
\volumenumber{ }
\issuenumber{ }
\eid{ }
\date{\today}
\startpage{1}
\endpage{102}
\maketitle

Recent progress in the trapping of cold atoms has attracted great interest
related to a wide variety of fields: condensed matter physics, atomic,
molecular and optical physics, astrophysics and physics of quark and nuclear
matter. There exists a deep analogy between the dynamics of cold atoms,
astrophysical systems, nuclear and plasma systems
\cite{Casalbuoni2004,Litvinov,Braaten,Labeyrie}. The possibility to tune the
effective interaction strength and to control populations of different spin
states provides a unique opportunity to investigate various phenomena of
interacting many-body systems. Imbalanced fermi superfluids are in particular
relevant for color superconductivity in dense quark matter \cite{quark,Alford}
and for neutron-proton pairing in asymmetric nuclear matter \cite{nuclear}.

In experiments \cite{K2,Z2006,Ketterle2006,Partridge2006,P2,P3}, a phase
separation between the superfluid and normal component of an interacting Fermi
gas of cold atoms with unequal spin populations has been observed. These
experiments identify a shell structure of the fermion cloud, in which a
superfluid core is surrounded with a Fermi gas in the normal state.

The mean-field approach provides a convincing qualitative explanation of the
phase separation of imbalanced cold fermion atoms in a trap
\cite{DeSilva,Haque,Haque2}. Mean field predicts a superfluid core (which at
$T=0$ is completely unpolarized) surrounded with an imbalanced normal phase,
and a discontinuous behavior of the majority and minority component densities
at the phase boundary. The theoretical study of a trapped strongly interacting
Fermi gas in the unitarity limit at the zero temperature \cite{Chevy2006}
shows a good agreement of the radii for the majority and minority components
with the experimental results of Ref. \cite{Partridge2006}. However, by
contrast with experiments, the boundary between the two phases provided by the
mean-field approach (see Refs. \cite{DeSilva,Haque,Haque2,Chevy2006} and the
mean-field results below in the present work) is very sharp both for elongated
and oblate traps, i. e., independently of a trap anisotropy, while
experimentally observed phase boundaries are rather smoothed in the case of
Ref. \cite{Ketterle2006}. Here, we describe quantitatively this smoothing by
taking into account effects beyond the mean-field approximation, at nonzero
temperature. The density profiles derived in this work offer the possibility
to determine the temperature of the superfluid by a fitting procedure, all
other parameters being fixed by the experimental configuration.

Recent experiments investigating pairing of cold atoms are realized in the
crossover regime between weak coupling, where the Bardeen-Cooper-Schrieffer
(BCS) theory holds, and strong coupling where the system is well described as
a Bose-Einstein condensate (BEC) of molecules. In the BCS-BEC crossover
regime, the critical temperature $T_{c}$ of the superfluid phase transition is
substantially lowered with respect to $T_{c}$ obtained within the mean-field
approach. The lowering of $T_{c}$ is due to Gaussian fluctuations about the
saddle point. The fluctuations influence also other parameters of the fermion
system, such as the density. In addition to fluctuations, also other factors
influence the density profiles calculated for cold Fermi gases: the possible
violation of the local density approximation (LDA) in highly anisotropic traps
observed by Hulet and co-workers \cite{Partridge2006,P2,P3} or the occurrence
of nontrivial phases such as Fulde-Ferrell-Larkin-Ovchinnikov state as
reported by Yoshida and Yip \cite{Yoshida2007}. These mechanisms can be
described using the method outlined below, but this is beyond the scope of the
present treatment, which focuses on the role of fluctuations. High-resolution
density distributions of cold $^{6}$Li atoms measured by Ketterle and
co-workers \cite{Ketterle2006} allow us to perform a quantitative comparison
between the experiment and theory.

There are different techniques to treat degenerate Fermi gases in the BCS-BEC
crossover regime. The diagrammatic technique is applied by Perali et al.
\cite{strinati}, Chen et al. \cite{levin}, and Taylor et al. \cite{Taylor2006}
to extend the results obtained by Nozi\`{e}res and Schmitt-Rink (NSR)
\cite{NSR} for $T=T_{c}$ to arbitrary temperatures. The NSR-like scheme
\cite{Hu2006} provides the equation of state which is in an excellent
agreement with Monte Carlo calculations within the whole range of the BCS-BEC
crossover at finite temperatures. This approach is also capable to show the
universal thermodynamics of strong interacting fermions \cite{Hu2007,Hu2008}.

In order to investigate the phase separation of an interacting Fermi gas in a
trap, we use the path-integral formalism
\cite{demelo1993,Engelbrecht1997,Diener2008}. The path-integral method is
attractive because it allows to obtain reliable parameters for interacting
fermions in the whole range of the coupling strength. However, it is necessary
to go beyond the low-temperature approximation of Refs.
\cite{Engelbrecht1997,Diener2008}, because the contribution to the density due
to fluctuations is sensitive to the temperature variations, even in the
low-temperature region.

In the present work, the path-integral approach is extended to the case of
arbitrary finite temperature and imbalanced spin populations, and applied to
calculate fluctuation contributions to the density. We consider a
two-component Fermi gas with the spin states $\left\vert +\right\rangle $ and
$\left\vert -\right\rangle $ in an anisotropic parabolic trap within the LDA.
The coordinate dependence of the density of fermions in a given spin state is
determined through the chemical potential
\begin{equation}
\mu_{\pm}\left(  \mathbf{r}\right)  =\mu_{\pm}\left(  0\right)  -\frac{m}%
{2}\left[  \omega_{1}^{2}\left(  x^{2}+y^{2}\right)  +\omega_{2}^{2}%
z^{2}\right]  \label{mu}%
\end{equation}
characterized by the confinement frequencies $\omega_{1}$ and $\omega_{2}$.
For given values of the local density $n=n_{+}+n_{-}$ and of the local density
difference $\delta n=n_{+}-n_{-}$, the averaged chemical potential $\mu
\equiv\left(  \mu_{+}+\mu_{-}\right)  /2$, the imbalance potential
$\zeta\equiv\left(  \mu_{+}-\mu_{-}\right)  /2$ and the gap parameter $\Delta$
are found as a joint solution of the gap equation and of the two number
equations. The thermodynamic potential and therefore the densities are
determined in the quadratic approximation with respect to the quantum
fluctuations about the saddle point for arbitrary temperatures and taking into
account the population imbalance.

Within the NSR scheme, the Gaussian fluctuations do not feed back into the
saddle-point equation for the functional integral, and the fluctuations only
contribute to the fermion density and to the number equation. Hence the gap
equation takes the same form as in the mean-field approximation. As shown in
Ref. \cite{Diener2008}, this is a natural approximation within the
path-integral formalism.

For a balanced gas, the saddle-point thermodynamic potential has a single
minimum, which results in the gap equation. For an imbalanced gas, however,
the saddle-point thermodynamic potential can have two minima at $\Delta=0$ and
at $\Delta\neq0$ \cite{Bedaque}. Therefore, the phase boundary for an
imbalanced Fermi gas in a trap cannot be determined from the gap equation, as
well as from the Thouless criterion \cite{Thouless1960}, because they both
allow one to find only a \emph{local} minimum of the variational problem for
the thermodynamic potential \cite{Thouless1960}. As a natural extension of the
NSR scheme to an imbalanced Fermi gas, we find coordinate-dependent values of
the gap parameter through the straightforward minimization of the saddle-point
thermodynamic potential \cite{JT2008},
\begin{align}
\frac{\Omega_{sp}}{V}  &  =-\int\frac{d\mathbf{k}}{\left(  2\pi\right)  ^{3}%
}\left[  \frac{1}{\beta}\ln\left(  2\cosh\beta\zeta+2\cosh\beta E_{\mathbf{k}%
}\right)  -\xi_{\mathbf{k}}-\frac{\left\vert \Delta\right\vert ^{2}}{2k^{2}%
}\right] \nonumber\\
&  -\frac{\left\vert \Delta\right\vert ^{2}}{8\pi a_{s}}, \label{Fsp}%
\end{align}
where $a_{s}$ is the scattering length, $\xi_{\mathbf{k}}=k^{2}-\mu$ is the
fermion energy, and $E_{\mathbf{k}}=\sqrt{\xi_{\mathbf{k}}^{2}+\left\vert
\Delta\right\vert ^{2}}$ is the Bogolubov excitation energy. We use the units
in which $\hbar=1$, $m=1/2$, and the Fermi energy $E_{F}=1$.

The fermion densities are a sum of mean-field and fluctuation contributions.
The current finite temperature implementation of the extended NSR scheme is
similar to that used in studying $d$-wave pairing in Ref. \cite{JT2008}. For
the case of the $s$-wave pairing mechanism, we find that the fluctuation
contribution $n_{fl}$ to the total density $n$ is given by%
\begin{align}
n_{fl}  &  =-\int\frac{d\mathbf{q}}{\left(  2\pi\right)  ^{3}}\left[  \frac
{1}{\pi}\int_{-\infty}^{\infty}\operatorname{Im}\frac{J\left(  \mathbf{q}%
,\omega+i\gamma\right)  }{e^{\beta\left(  \omega+i\gamma\right)  }-1}%
d\omega\right. \nonumber\\
&  \left.  +\frac{1}{\beta}\sum_{n=-n_{0}}^{n_{0}}J\left(  \mathbf{q}%
,i\Omega_{n}\right)  \right]  , \label{nfl}%
\end{align}
with $\beta$ the inverse to the temperature. Here, $n_{0}$ is an arbitrary
positive integer, and the parameter $\gamma$ lies between two bosonic
Matsubara frequencies $\Omega_{n_{0}}<\gamma<\Omega_{n_{0}+1}$, $\Omega
_{n}\equiv2\pi n/\beta$. For the computation, a fast convergence of the
integral over $\omega$ in (\ref{nfl}) is achieved when $\gamma=\left(
2n+1\right)  \pi/\beta$, and an integer $n$ is chosen in such a way that
$\gamma\sim1$. The function $J\left(  \mathbf{q},z\right)  $ of complex
argument $z$ is%
\begin{equation}
J\left(  \mathbf{q},z\right)  =\frac{M_{1,1}\left(  \mathbf{q},-z\right)
\frac{\partial M_{1,1}\left(  \mathbf{q},z\right)  }{\partial\mu}%
-M_{1,2}\left(  \mathbf{q},-z\right)  \frac{\partial M_{1,2}\left(
\mathbf{q},z\right)  }{\partial\mu}}{M_{1,1}\left(  \mathbf{q},z\right)
M_{1,1}\left(  \mathbf{q},-z\right)  -M_{1,2}^{2}\left(  \mathbf{q},z\right)
}. \label{FJ}%
\end{equation}
The fluctuation contribution $\delta n_{fl}$ to the density difference $\delta
n$ is given by the same expression as (\ref{nfl}) with replacing the
derivatives $\partial/\partial\mu$ in the function $J\left(  \mathbf{q}%
,z\right)  $ by $\partial/\partial\zeta$. The matrix elements $M_{j,k}\left(
\mathbf{q},z\right)  $ are given by
\begin{align}
M_{1,1}\left(  \mathbf{q},z\right)   &  =\int\frac{d\mathbf{k}}{\left(
2\pi\right)  ^{3}}\left[  \frac{1}{2k^{2}}+\frac{\sinh\beta E_{\mathbf{k}}%
}{2E_{\mathbf{k}}\left(  \cosh\beta E_{\mathbf{k}}+\cosh\beta\zeta\right)
}\right. \nonumber\\
&  \times\left(  \frac{\left(  z-E_{\mathbf{k}}+\varepsilon_{\mathbf{k}%
+\mathbf{q}}\right)  \left(  E_{\mathbf{k}}+\varepsilon_{\mathbf{k}}\right)
}{\left(  z-E_{\mathbf{k}}+E_{\mathbf{k}+\mathbf{q}}\right)  \left(
z-E_{\mathbf{k}}-E_{\mathbf{k}+\mathbf{q}}\right)  }\right. \nonumber\\
&  \left.  \left.  -\frac{\left(  z+E_{\mathbf{k}}+\varepsilon_{\mathbf{k}%
+\mathbf{q}}\right)  \left(  E_{\mathbf{k}}-\varepsilon_{\mathbf{k}}\right)
}{\left(  z+E_{\mathbf{k}}-E_{\mathbf{k}+\mathbf{q}}\right)  \left(
z+E_{\mathbf{k}+\mathbf{q}}+E_{\mathbf{k}}\right)  }\right)  \right]
\nonumber\\
&  -\frac{1}{8\pi a_{s}}, \label{M11}%
\end{align}%
\begin{align}
M_{1,2}\left(  \mathbf{q},z\right)   &  =-\left\vert \Delta\right\vert
^{2}\int\frac{d\mathbf{k}}{\left(  2\pi\right)  ^{3}}\frac{\sinh\beta
E_{\mathbf{k}}}{2E_{\mathbf{k}}\left(  \cosh\beta E_{\mathbf{k}}+\cosh
\beta\zeta\right)  }\nonumber\\
&  \times\left(  \frac{1}{\left(  z-E_{\mathbf{k}}+E_{\mathbf{k}+\mathbf{q}%
}\right)  \left(  z-E_{\mathbf{k}}-E_{\mathbf{k}+\mathbf{q}}\right)  }\right.
\nonumber\\
&  \left.  +\frac{1}{\left(  z+E_{\mathbf{k}}-E_{\mathbf{k}+\mathbf{q}%
}\right)  \left(  z+E_{\mathbf{k}}+E_{\mathbf{k}+\mathbf{q}}\right)  }\right)
, \label{M22}%
\end{align}
where $a_{s}$ is the $s$-wave scattering length and $E_{\mathbf{k}}%
=\sqrt{\varepsilon_{\mathbf{k}}^{2}+\Delta^{2}}$ is the energy of the
Bogolubov excitation with $\varepsilon_{\mathbf{k}}=k^{2}-\mu$. Here, we use
the units with the mass $m=1/2$, $\hbar=1$, and the Fermi energy $E_{F}=1$.
These expressions incorporate not only the particle-pair and hole-pair
excitations, but also particle-hole excitations, so that in the limit
$\Delta\rightarrow0$ one obtains an interacting Fermi gas rather than the
ideal gas.

The phase transitions of the imbalanced fermion gas were analyzed using phase
diagrams for various thermodynamic variables (see Refs.
\cite{Koponen2007,Zhang2007,Parish2007,Pao,Sheehy}). The finite-temperature
phase diagram pressure/temperature \cite{Chien,He} is probably the first phase
diagram which includes finite temperature and fluctuation effects as well as
trap inhomogeneity. At a fixed scattering length $a_{s}$, a temperature
$T_{0}$ of a tricritical point separates two types of the phase transition as
follows. For $T\geq T_{0}$, the relative population imbalance $\delta n/n$ is
continuous across the phase transition, so that the latter one is of the
second order. On the contrary, at $T<T_{0}$, the relative population imbalance
changes discontinuously at the phase transition, so that the latter one os of
the first order \cite{Bedaque}. As a result, there exists the region where a
uniform normal or superfluid state cannot exist, and therefore in this region
phase separation occurs \cite{Pao,Sheehy,Parish2007}.

The effect of fluctuations beyond mean-field on the phase diagram is not
completely resolved \cite{Chien,He}, although the effect of Gaussian
fluctuations on the locus of the tricritical points was studied in
\cite{Parish2007}. In this contribution, rather than focusing on the hard task
of correcting the mean-field phase diagram, we will investigate the effects of
fluctuations on the density profiles, where the fluctuation contribution is
straightforwardly obtained from Eq. (\ref{nfl}).

The phase separation exists at temperatures lower than the temperature of the
tricritical point. Typical temperatures of the experiments on the phase
separation of an imbalanced mixture of cold atoms are estimated to be as low
as $T\sim0.1T_{F}$, where $T_{F}$ is the Fermi temperature $T_{F}=E_{F}/k_{B}%
$. Therefore, the phase transition at the boundary of the superfluid core in
the experiments \cite{Ketterle2006,Partridge2006} is expected to be of the
first order. In order to get a definite answer to this question, we
investigate the distributions of the gap parameter and of the density for a
fermion gas in an anisotropic parabolic trap taking into account Gaussian
fluctuations and using the parameters typical for the conditions in the
experiment by Ketterle and co-workers \cite{Ketterle2006}.%

\begin{figure}
[h]
\begin{center}
\includegraphics[
height=4.1978in,
width=3.1695in
]%
{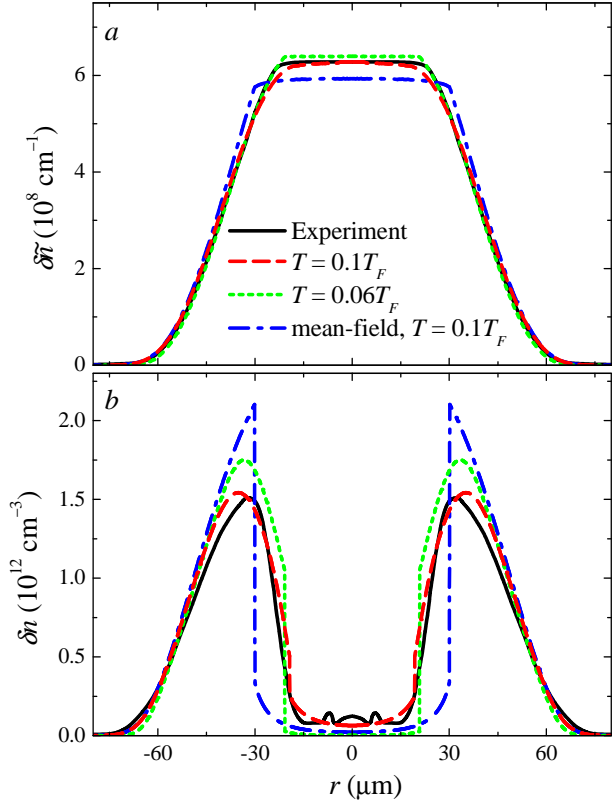}%
\caption{(color online). (\emph{a}) The distribution profiles for the
integrated density difference compared with the experimental data (black
curve) from Fig. 3(d) of Ref. \cite{Ketterle2006}. (\emph{b}) The 3D
distribution profiles along the lateral direction for the local density
difference compared with the experimental 3D distribution profile (black
curve) from Fig. 3(e) of Ref. \cite{Ketterle2006}. Dashed and dotted curves
correspond to the present approach taking into account Gaussian fluctuations,
for $T=0.1T_{F}$ and $T=0.06T_{F}$, respectively. Dot-dashed curves correspond
to the mean-field approach.}%
\label{fig:profiles}%
\end{center}
\end{figure}

In Fig. \ref{fig:profiles}~(\emph{a}), we plot 1D profiles $\delta\tilde{n}$
obtained by the integration of $\delta n$ over $y$ and $z$ coordinates
(corresponding to the profile of the integrated density difference from
Fig.~3(d) of Ref. \cite{Ketterle2006}). Fig. \ref{fig:profiles}~(\emph{b})
shows the 3D distribution profiles along the lateral direction for the density
difference, compared with the experimental distribution profiles from Ref.
\cite{Ketterle2006}. Both with and without fluctuations, there is a spatial
phase separation, where a superfluid core is surrounded by a normal phase. In
the trap at a sufficiently low temperature the fermion system is almost
exactly balanced in the core and almost completely polarized outside the core.
Local densities for both majority and minority components change their values
sharply when passing the phase boundary. Correspondingly, the total fermion
density and the local density difference reveal discontinuities at the phase
boundary. Those discontinuities are a consequence of the fact that the phase
transition is of the first order.

Within the BCS-BEC crossover regime, the Gaussian fluctuations lead to a
substantial decrease of the core radius with respect to that obtained in the
mean-field approximation. The fluctuations reduce the amplitude of the
discontinuous change of the density difference and provide a partially
smoothed (with a relatively small discontinuity) density profile near the
phase boundary. A smoothing of the radial density profiles owing to
fluctuations was also obtained by Chien \emph{et al}. \cite{Chien,He}. The
profile for $\delta n,$ obtained taking into account fluctuations, lies
drastically closer to the experimental radial profile \cite{Ketterle2006} for
the local population imbalance than the mean-field result. As well as the
radial distributions, the 1D profiles of the integrated density difference
calculated taking into account Gaussian fluctuations are in a better agreement
with the experiment \cite{Ketterle2006} than the mean-field results in what
regards the core size and the density profile. With respect to the present
approach and to the experiment, the mean-field approximation overestimates the
size of the superfluid core and the magnitude of the jump of the density
difference at the phase boundary.%

\begin{figure}
[h]
\begin{center}
\includegraphics[
height=2.2554in,
width=3.1462in
]%
{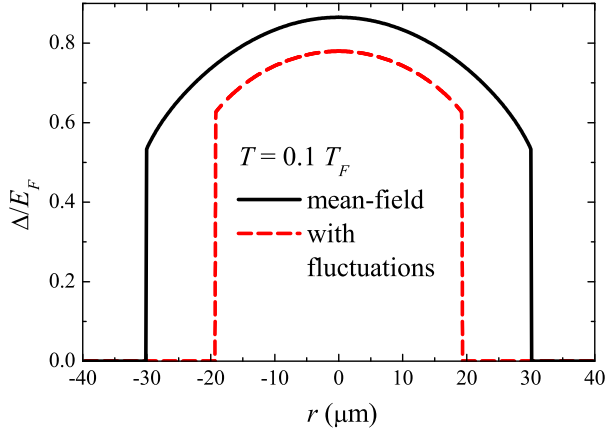}%
\caption{(color online) 3D distribution profiles along the lateral direction
for the gap parameter calculated without (solid curve) and with (dashed curve)
Gaussian fluctuations for cold fermions in the parabolic trap at $T=0.1T_{F}$
and $\delta N/N=0.58$.}%
\label{fig:order}%
\end{center}
\end{figure}

The 3D distribution profiles along the lateral direction for the gap parameter
calculated within the mean-field approximation and taking into account
Gaussian fluctuations are plotted in Fig. \ref{fig:order}. As seen from the
comparison of the spatial distributions of the gap parameter obtained within
the mean-field approach and within the NSR scheme, the fluctuations reduce
both the core radius and the values of the gap parameter inside the core. At
the phase boundary between the superfluid and normal phases, the gap parameter
discontinuously changes its magnitude. When taking fluctuations into account,
the first-order phase transition between the superfluid and normal states
occurs at a higher magnitude of the gap parameter than within the mean-field
approach. This is a consequence of the fact that fluctuations lower the
stability of the superfluid state.

As found in Refs. \cite{strinati,Taylor2006,Parish2007,Chien,He}, the NSR
scheme leads to difficulties near the unitarity limit, where the fluctuation
contribution to the fermion density is not small and therefore can be hardly
treated as a perturbation. For example, close to the critical temperature, the
NSR approach predicts a bend-over behavior of the gap parameter as a function
of the temperature \cite{Taylor2006}. Parish \emph{et al}. \cite{Parish2007}
showed that near unitarity, the susceptibility matrix%
\begin{equation}
\left\Vert \chi_{jk}\right\Vert =\left(
\begin{array}
[c]{cc}%
\left(  \frac{\partial n_{+}}{\partial\mu_{+}}\right)  _{T,\Delta,\mu_{-}} &
\left(  \frac{\partial n_{+}}{\partial\mu_{-}}\right)  _{T,\Delta,\mu_{+}}\\
\left(  \frac{\partial n_{-}}{\partial\mu_{+}}\right)  _{T,\Delta,\mu_{-}} &
\left(  \frac{\partial n_{-}}{\partial\mu_{-}}\right)  _{T,\Delta,\mu_{+}}%
\end{array}
\right)  \label{comp}%
\end{equation}
within the NSR scheme is not positive semi-definite. They interpreted this
result as a breakdown of the NSR treatment. However, this breakdown occurs
when the fluctuation contributions are sufficiently high -- at temperatures
close to $T_{c}$ for a uniform Fermi gas. Also the bend-over behavior of the
gap parameter predicted by Taylor \emph{et al}. \cite{Taylor2006} occurs near
the critical temperature for a balanced uniform Fermi gas $T_{c}%
\approx0.23T_{F}$ \cite{demelo1993}.

The fluctuation contribution to the fermion density strongly falls down with
decreasing temperature. The experiments of the MIT group
\cite{Ketterle2006,Partridge2006} are performed for lower temperatures
$T\lesssim0.1T_{F}$, at which the fluctuation contribution to the density is
smaller than at $T_{c}$ for a balanced uniform gas. In this connection, we can
expect that the fluctuation contribution to the fermion density is relatively
small with respect to the mean-field contribution, and therefore the extended
NSR scheme can be applicable to the experimental conditions of Refs.
\cite{K2,Z2006,Ketterle2006,Partridge2006,P2,P3}. To verify this assumption,
we study the behavior of the susceptibility matrix for cold fermions in a trap.

The positive semi-definite matrix means that its eigenvalues are non-negative.
The eigenvalues $\lambda_{1}$ and $\lambda_{2}$ of the matrix (\ref{comp}) can
be expressed through the derivatives of the total density $n$ and of the
density difference $\delta n$ using as the independent variables the averaged
chemical potential $\mu$ and the imbalance potential $\zeta$:%
\begin{align}
\lambda_{1,2}  &  =\frac{1}{4}\left\{  \left(  \frac{\partial n}{\partial\mu
}\right)  _{T,\Delta,\zeta}+\left(  \frac{\partial\left(  \delta n\right)
}{\partial\zeta}\right)  _{T,\Delta,\mu}\right. \nonumber\\
&  \mp\left.  \sqrt{\left[  \left(  \frac{\partial n}{\partial\mu}\right)
_{T,\Delta,\zeta}-\left(  \frac{\partial\left(  \delta n\right)  }%
{\partial\zeta}\right)  _{T,\Delta,\mu}\right]  ^{2}+4\left(  \frac{\partial
n}{\partial\zeta}\right)  _{T,\Delta,\mu}^{2}}\right\}  . \label{eigenv}%
\end{align}
As follows from the relation%
\begin{equation}
\left(  \frac{\partial n}{\partial\zeta}\right)  _{T,\Delta,\mu}=\left(
\frac{\partial\left(  \delta n\right)  }{\partial\mu}\right)  _{T,\Delta
,\zeta}, \label{rr}%
\end{equation}
in the limit of a vanishing imbalance, when $\delta n=0$, $\left(
\frac{\partial n}{\partial\zeta}\right)  _{T,\Delta,\mu}=0$. Therefore in this
case the eigenvalues of the susceptibility matrix are reduced to the
derivatives $\left(  \frac{\partial n}{\partial\mu}\right)  _{T,\Delta,\zeta}$
and $\left(  \frac{\partial\left(  \delta n\right)  }{\partial\zeta}\right)
_{T,\Delta,\mu}$.

\begin{figure}
[h]
\begin{center}
\includegraphics[
height=2.3056in,
width=3.1652in
]%
{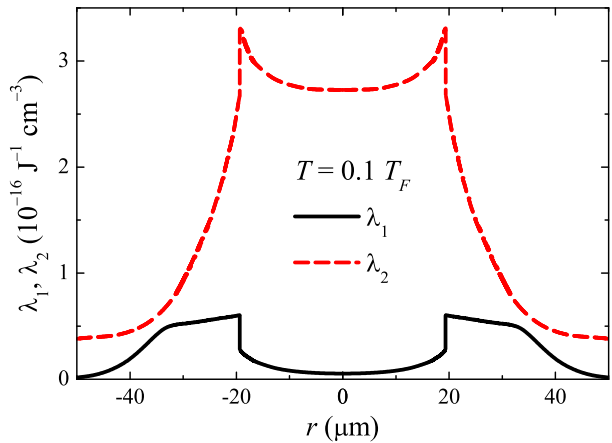}%
\caption{(color online) 3D distribution profiles for the eigenvalues
$\lambda_{1},\lambda_{2}$ of the susceptibility matrix for imbalanced
interacting fermions in the parabolic trap at $T=0.1T_{F}$ and $\delta
N/N=0.58$.}%
\label{fig:susc}%
\end{center}
\end{figure}

In order to check whether the susceptibility matrix is positive semi-definite
at the conditions of the experiments under consideration
\cite{Ketterle2006,Partridge2006}, we analyze the eigenvalues of the
susceptibility matrix as a function of the radius using the values of
parameters relevant to the experiment \cite{Ketterle2006}. The eigenvalues are
calculated taking into account Gaussian fluctuations about the saddle point.
The 3D distribution profiles for the eigenvalues of the susceptibility matrix
(\ref{eigenv}) are shown in Fig. \ref{fig:susc}.%

The susceptibility matrix eigenvalues, as well as the gap parameter and the
density profiles, behave discontinuously at the phase boundary. We see that in
the stable superfluid and normal states, both $\lambda_{1}$ and $\lambda_{2}$
are positive. This result shows that for the experimental conditions of Refs.
\cite{Ketterle2006,Partridge2006}, the NSR scheme extended to the imbalanced
case provides a reasonable interpretation of the phase separation of a trapped
Fermi gas.

Currently, the temperature in the experiment is determined by fitting to the
tails of the remnant thermal distributions. One of the implicit assumptions is
a good thermalization between the superfluid and the thermal cloud, and this
assumption has recently been under further investigation \cite{bert}. Owing to
fluctuations, the density profile is rather sensitive to temperature
variations at low temperatures $T\ll T_{F}$. The good agreement between the
temperature-dependent density profiles calculated here and the experimental
density profiles allows us to estimate the temperature of the superfluid core.

In conclusion, we have investigated the density profiles of a phase separated
imbalanced Fermi gas in an anisotropic parabolic trap, taking into account
Gaussian fluctuations near the saddle point at a nonzero temperature.
Fluctuations lead to a smoothing of the density profiles and to a decrease of
the radius of the superfluid core with respect to those predicted by the
mean-field approximation. The calculated density profiles are in a good
agreement with the experimental density profiles observed in Ref.
\cite{Ketterle2006} and their temperature dependence enables temperature
determination. The present approach allows us for a much more convincing
interpretation of the experimental results on the phase separation of
imbalanced fermions with respect to the mean-field approximation.

\begin{acknowledgments}
We are grateful to W.~Ketterle, M.~Wouters, D.~Lemmens, J.~O.~Indekeu,
M.~K.~Oberthaler, E.~Timmermans and H.~T.~C.~Stoof for valuable discussions.
This work has been supported by the FWO-V Project Nos. G.0356.06, G.0115.06,
G.0435.03, G.0306.00, the WOG Project No. WO.025.99N, and the NOI BOF UA 2004.
\end{acknowledgments}

\end{document}